\documentclass[a4paper, 12pt]{article}
\usepackage{fullpage}
\usepackage{amsmath, amsthm, amssymb, mathrsfs, graphicx}
\usepackage{ifthen, subfigure}
\usepackage[usenames]{color}
\usepackage{url}

\usepackage{natbib}
\bibpunct{(}{)}{;}{a}{}{,}

\theoremstyle{plain}

\theoremstyle{remark}

\theoremstyle{definition}

\theoremstyle{remark}
\newtheorem{ex}{Example}
\newtheorem{scenario}{Scenario}

\theoremstyle{definition}
\newtheorem*{PRalg}{PR algorithm}
\newtheorem*{PREMalg}{PR--EM algorithm}

\newcommand{\U}{\mathscr{U}}

\newcommand{\RR}{\mathbb{R}}

\newcommand{\nm}{\mathsf{N}}
\newcommand{\stt}{\mathsf{t}}

\newcommand{\eps}{\varepsilon}
\renewcommand{\phi}{\varphi}

\newcommand{\Lpr}{L_{\text{\sc pr}}}
\newcommand{\lpr}{\ell_{\text{\sc pr}}}

\newcommand{\Umin}{U_{\text{min}}}
\newcommand{\Umax}{U_{\text{max}}}

\newcommand{\xvec}{\boldsymbol{x}}
\newcommand{\yvec}{\boldsymbol{y}}

\newcommand{\vbeta}{\boldsymbol{\beta}}
\newcommand{\veps}{\boldsymbol{\varepsilon}}

\newcommand{\vtheta}{\boldsymbol{\theta}}

\newcommand{\vOmega}{\boldsymbol{\Omega}}

\newcommand{\Xmat}{\mathbf{X}}

\begin{document}

\title{A semiparametric scale-mixture regression model and predictive recursion maximum likelihood}
\author{Ryan Martin \\ Department of Mathematics, Statistics, and Computer Science \\ University of Illinois at Chicago \\ \url{rgmartin@uic.edu} \\
\mbox{} \\
Zhen Han \\ Department of Statistics \\ North Carolina State University \\ \url{zhan3@ncsu.edu}}
\date{\today}

\maketitle

\begin{abstract}
To avoid specification of the error distribution in a regression model, we propose a general nonparametric scale mixture model for the error distribution.  For fitting such mixtures, the predictive recursion method is a simple and computationally efficient alternative to existing methods.  We define a predictive recursion-based marginal likelihood function, and estimation of the regression parameters proceeds by maximizing this function.  A hybrid predictive recursion--EM algorithm is proposed for this purpose.  The method's performance is compared with that of existing methods in simulations and real data analyses.    

\medskip

\end{abstract}

\section{Introduction}
\label{S:intro}

Consider the standard linear regression model,
\[ \yvec = \Xmat \vbeta + \veps, \]
where $\yvec=(y_1,\ldots,y_n)^\top$ is a $n \times 1$ vector of response variables, $\Xmat$ is a $n \times p$ matrix of predictor variables, with $i$th row $\xvec_i=(x_{i1},\ldots,x_{ip})$, $\vbeta$ is a $p \times 1$ vector of regression coefficients, and $\veps=(\eps_1,\ldots,\eps_n)^\top$ is an $n \times 1$ vector of independent and identically distributed (iid) errors with common density $f$.  In classical linear model applications, one assumes that $f$ is a normal distribution with mean zero and unknown variance $\sigma^2$.  In this case, the ordinary least squares method provides the optimal estimates of $(\vbeta,\sigma^2)$.  However, if $f$ happens to be non-normal, in particular, if $f$ has heavier-than-normal tails, then the accuracy of the ordinary least squares solutions is lost.  

When the error density $f$ may be non-normal, one might consider an alternative to the normal model and ordinary least squares.  Model-free alternatives based on M-estimation \citep[e.g.,][]{huber1973, huber1981} include methods based on minimizing an objective function different from the sum of squared residuals, such as least absolute deviation, or $L_1$ regression.  Surveys of these standard techniques are given in \citet{rousseeuw.leroy.1987} and \citet{ryan2009}.  If a likelihood-based method is preferred, then one common approach is to model the errors by a heavy-tailed Student-t distribution; see, for example, \citet{lange.little.taylor.1989}, \citet{cliu.1996}, and \citet{pinheiro.liu.wu.2001}.  The standard implementation of this approach uses the expectation--maximization (EM) algorithm \citep{dlr}, which is based on a representation of the Student-t distribution as a scale mixture of normals \citep{andrews.mallows.1974, west1987}.  The goal of this paper is to explore a more general version of this latter heavy-tailed model.  

Motivated by the Student-t's normal scale mixture representation, we consider a more general regression model specified by an arbitrary normal scale mixture for the error distribution.  Specifically, we write the error density $f$ as a mixture 
\begin{equation}
\label{eq:mixture}
f(\eps) = \int_0^\infty \nm(\eps \mid 0, u^2) \, \Psi(du), 
\end{equation}
where $\Psi$ is an unspecified mixing distribution supported on $(0,\infty)$.  By symmetry of the normal kernel, the density $f$ is symmetric.  Moreover, \eqref{eq:mixture} contains both the  normal model, $\nm(0,\sigma^2)$ and the Student-t model, $\stt_\nu(0, \sigma)$, as special cases, corresponding to $\Psi$ a point-mass at $\sigma$ and a scaled inverse chi-square distribution, respectively.  Since $\Psi$ is completely unspecified, an additional scale parameter would not be identifiable so, without loss of generality, $\Psi$ fully characterizes the error distribution in our regression model.  

To fit this new semiparametric regression model, estimation of both $\vbeta$ and $\Psi$ is required and, even though the mixing distribution $\Psi$ is a nuisance parameter, care is needed.  Maximum likelihood and Bayes approaches can be developed, and we discuss the computational challenges faced by these in Section~\ref{SS:challenges}.  The main contribution of this paper is a computationally efficient alternative, an extension of the predictive recursion (PR) method discussed in \citet{nqz}, \citet{newton02}, \citet{ghoshtokdar}, \citet{martinghosh}, \citet{tmg}, and \citet{mt-rate}.  The PR algorithm was originally designed for fast nonparametric estimation of a mixing distribution of a mixture model, but \citet{mt-prml} developed a PR-driven marginal likelihood approach for estimating structural parameters in semiparametric mixture models; see Section~\ref{SS:pr} for a brief review of the PR algorithm and related methods.  Previous applications of PR focused on location mixtures, and the special scale mixture formulation in this paper requires new ideas.  After writing down the PR marginal likelihood for the semiparametric regression problem, in Section~\ref{SS:em} we propose a hybrid PR--EM strategy that takes advantage of the latent scale parameter structure in the mixture model \eqref{eq:mixture}.  This hybrid algorithm is fast and easy to compute, and in Section~\ref{SS:ascent} we provide some theoretical support for its ascent property.  Some remarks on the robustness of the PR method are given in Section~\ref{SS:robustness}.  Section~\ref{S:numerical} demonstrates numerically that our proposed approach provides accurate estimation of $\vbeta$ compared to existing methods across a range of different error distributions.  Section~\ref{S:discuss} provides some concluding remarks.

\section{Background}
\label{S:background}

\subsection{Challenges faced by standard approaches}
\label{SS:challenges}

\begin{figure}
\begin{center}
\scalebox{0.5}{\includegraphics{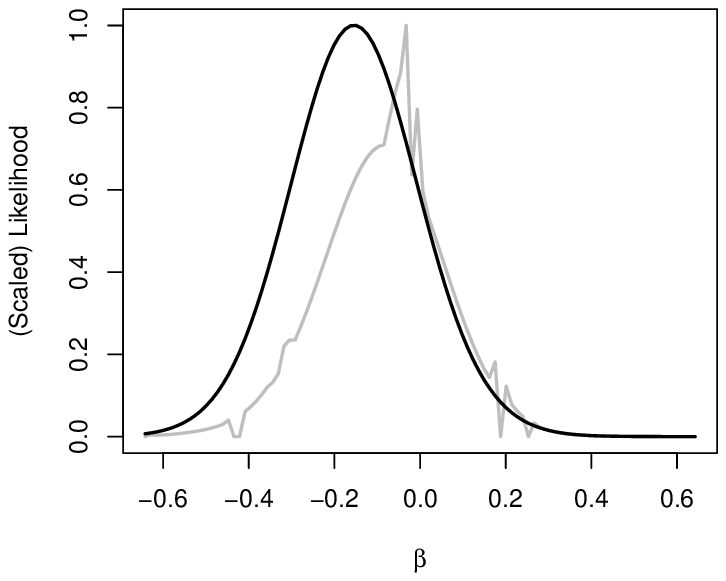}}
\end{center}
\caption{Plots of the (scaled) likelihood functions: profile likelihood based on nonparametric maximum likelihood (gray) and PR marginal likelihood (black).}
\label{fig:likpic}
\end{figure}

There are two natural likelihood-based approaches that one could consider for fitting the semiparametric regression model with error distribution \eqref{eq:mixture}.  The first is via nonparametric maximum likelihood.  Start by writing a joint likelihood function for $(\vbeta, \Psi)$:
\[ L(\vbeta, \Psi) = \prod_{i=1}^n \int_0^\infty \nm(y_i - \xvec_i \vbeta \mid 0, u^2) \Psi(du). \]
Next, define a profile likelihood $L_p(\vbeta) = L(\vbeta, \hat\Psi_{\vbeta})$, where $\hat\Psi_{\vbeta}$ is the conditional maximum likelihood estimator of $\Psi$ for the given $\vbeta$.  Then $L_p(\vbeta)$ can be treated like a usual likelihood function, to produce estimators, tests, or confidence regions for $\vbeta$.  Existing algorithms for nonparametric maximum likelihood estimation of mixing distributions \citep[e.g.,][]{wang} can be used to compute $\hat\Psi_{\vbeta}$ and, in turn, the profile likelihood $L_p(\vbeta)$.  We claim that this profile likelihood function is generally rough, so optimization over $\vbeta$ is unstable and computationally expensive.  To justify this claim, we consider a simple special case of the regression problem with no predictor variables, i.e., iid data with location $\beta$.  In this case, using Wang's algorithm, we can easily evaluate and plot the profile likelihood function.  An independent sample of size $n=100$ was drawn from a Student-t distribution with $\text{df}=2$, centered at $\beta=0$, and the corresponding likelihood functions for $\beta$ are plotted in Figure~\ref{fig:likpic}.  The profile likelihood has a number of local modes, so numerical optimization is unstable.  On the other hand, the likelihood function for our proposed method, described in Section~\ref{S:prml.reg}, is smooth with one global mode, so optimization is fast and easy.  

A second approach is based on nonparametric Bayes, where a prior distribution for $\Psi$ is introduced.  A reasonable choice would be to take a Dirichlet process prior for $\Psi$ \citep[e.g.,][]{ferguson1973, lo1984, muller.quintana.2004}.  The idea is to integrate out $\Psi$ from the joint likelihood $L(\vbeta,\Psi)$ with respect to the prior, leaving a marginal likelihood function $L_m(\vbeta)$ for $\vbeta$.  Markov chain Monte Carlo algorithms \citep[e.g.,][]{escobar.west.1995, maceachern.muller.1998, neal2000, carvalho.ba.2010}, as well as software \citep[e.g.,][]{dppackage}, are available for evaluating this marginal likelihood but this is too expensive because each marginal likelihood evaluation requires its own Monte Carlo run, and optimization requires several such runs.  One can avoid repeatedly running Monte Carlo if $\vbeta$ is assigned a proper prior.  That is, one can employ the technique in \citet{chib1995} to get a marginal likelihood for $\vbeta$ from a single, joint Monte Carlo run for $(\vbeta,\Psi)$.  This single joint Monte Carlo is generally more expensive than our proposed PR-driven strategy, so we do not explore this further here.

\subsection{Review of predictive recursion}
\label{SS:pr}

\subsubsection{Nonparametric mixtures}
\label{SSS:pr.non}

PR is a fast algorithm designed for recursive estimation of mixing distributions in nonparametric mixture models.  It was first proposed as an alternative to Markov chain Monte Carlo methods in fitting Bayesian Dirichlet process mixture models \citep{nqz, newton02}.  To summarize the general case, let $y_1,\ldots,y_n$ be iid with density $f(y)$, where $f=f_\psi$ is modeled as a mixture $\int_\U p(y \mid u) \psi(u) \,\lambda(du)$, $p(y \mid u)$ is a known kernel, and $\psi$ is an unknown density with respect to a dominating $\sigma$-finite measure $\lambda$ on $\U$.  The PR algorithm estimates $\psi$ and $f_\psi$ as follows.  

\begin{PRalg}
Initialize the algorithm by choosing a $\lambda$-density $\psi_0$ and a sequence of weights $\{w_i: i \geq 1\} \subset (0,1)$.  For $i=1,\ldots,n$, repeat the following steps.
\begin{enumerate}
\item Compute the mixture density: 
\begin{equation}
\label{eq:pr.mixture}
f_{i-1}(\cdot) \equiv f_{\psi_{i-1}}(\cdot) = \int_\U p(\cdot \mid u) \psi_{i-1}(u) \,\lambda(du).
\end{equation}
\item Update the mixing density estimate: 
\begin{equation}
\label{eq:pr.mixing}
\psi_i(u) = (1-w_i) \,\psi_{i-1}(u) + w_i \, \frac{p(y_i \mid u) \psi_{i-1}(u)}{f_{i-1}(y_i)}.  
\end{equation}
\end{enumerate}
Return $\psi_n$ and $f_n=f_{\psi_n}$ as the PR estimates of $\psi$ and $f_\psi$, respectively.  
\end{PRalg}

Two key properties of the PR algorithm are speed and ease of implementation.  Also, PR is able to produce an estimate of the mixing distribution which has a density with respect to the prescribed dominating measure $\lambda$.  Large-sample convergence properties of the PR estimates are given in \citet{tmg}.  In particular, under suitable conditions, the PR estimate $f_n$ of the mixture density is  consistent and, if the mixing distribution is identifiable, then the PR estimate $\psi_n$ of the mixing density is also consistent.  \citet{mt-rate} provides bounds on the PR rate of convergence.  

To end this subsection, we discuss a few specific properties of the PR algorithm that are relevant to its implementation.  
\begin{itemize}
\item The weights $\{w_i\}$ in the PR algorithm are required to satisfy $\sum_{i=1}^\infty w_i = \infty$ and $\sum_{i=1}^\infty w_i^2 < \infty$.  Subject to these conditions, the practical performance of PR is not too sensitive to the particular choice.  Here we take $w_i = (i+1)^{-1}$.  
\item The PR estimates depend on the order in which the data are processed.  This dependence can be weakened by averaging the PR estimates over several (random) permutations of the data sequence.  In our experience, averaging over 25 permutations is sufficient \citep{mt-test} and, given the speed of PR, this does not significantly increase the computational cost.  
\end{itemize}

\subsubsection{Semiparametric mixtures}

As an extension of the nonparametric mixture model setup in the previous subsection, consider the case where the kernel $p(y \mid u)$ depends on an unknown parameter $\vtheta$, i.e., $p(y \mid u) = p(y \mid \vtheta, u)$.  In this context, the structural parameter $\vtheta$ is typically of primary interest, while the mixing density $\psi$ is a nuisance parameter.  

For this problem, \citet{mt-prml} proposed an extension of the PR algorithm that produces a sort of likelihood function for $\vtheta$ to be used for inference.  Let $f_{k,\vtheta}$ be the PR mixture density estimate based on data $y_1,\ldots,y_k$, $k=1,\ldots,n$, and kernel $p(y \mid \vtheta, u)$, where $\vtheta$ is taken to be fixed.  Consider the function 
\[ \Lpr(\vtheta) = \prod_{i=1}^n f_{i-1,\vtheta}(y_i). \]
This function is called the PR marginal likelihood for $\vtheta$.  Despite its familiar product-of-densities form, this is not a genuine likelihood function for $\vtheta$ under the posited semiparametric model.  \citet{mt-prml} use PR's natural connection to the Bayesian Dirichlet process prior model to argue that $\Lpr(\vtheta)$ is an approximate marginal likelihood for $\vtheta$.  They also demonstrate the large-sample convergence properties of $\Lpr(\vtheta)$, and give some examples.  See, also, \citet{mt-test} and \citet{prml-finite}.

\section{PR maximum likelihood for regression}
\label{S:prml.reg}

\subsection{Formulation}
\label{SS:method}

Consider the linear model $\yvec = \Xmat\vbeta + \veps$, where $\veps$ is an $n$-vector of iid errors assumed to have density $f$ of the mixture form in \eqref{eq:mixture}, where the mixing density $\psi$, supported on $\U \subseteq (0,\infty)$ is unknown.  As discussed in Section~\ref{S:intro}, $f$ has heavier-than-normal tails, so inference on $\vbeta$ based on such a model will be less sensitive to extreme observations compared to inference based on a basic normal model.  To put this in the form suitable for PR, write the mixture model for the residuals,
\[ f(y_i-\xvec_i \vbeta) = \int_\U \nm(y_i - \xvec_i \vbeta \mid 0, u^2) \psi(u) \,du, \quad i=1,\ldots,n. \]
Then we can apply the PR algorithm to the residuals, $y_i - \xvec_i \vbeta$, to estimate the mixing density.  If $f_{k,\vbeta}$ is the PR estimate of the mixture density for the given $\vbeta$ based on $y_1,\ldots,y_k$, $k=1,\ldots,n$, then we get the following PR marginal likelihood for $\vbeta$:
\begin{equation}
\label{eq:prml.reg}
\Lpr(\vbeta) = \prod_{i=1}^n f_{i-1,\vbeta}(y_i - \xvec_i \vbeta). 
\end{equation}
This is fast and easy to compute.  As with all likelihood functions, we propose to estimate $\vbeta$ by maximizing this PR marginal likelihood or, equivalently, the PR log-marginal likelihood $\lpr(\vbeta) = \log \Lpr(\vbeta)$.  More on this in Section~\ref{SS:em}.  

There are three important remarks concerning implementation of this approach.  
\begin{itemize}
\item For PR computations, a compact support $\U \subset (0,\infty)$ is required; compact mixing distribution support was also a general suggestion made in \citet{rogers.tukey.1972}.  We take $\U = [\Umin, \Umax]$, where $\Umin$ is fixed at $10^{-5}$, and $\Umax$ is to be specified.  Since $\Umax$ helps to determine the overall scale of the error distribution, we should select $\Umax$ to satisfy two criteria.  First, $\Umax$ should be sufficiently large so that the support is not overly restricted.  Second, if the errors are actually normal with scale $\sigma$, then the PR method should be able to recover this by producing an estimate of $\psi$ that is tightly concentrated around the usual root mean square error estimator $\hat\sigma$ of $\sigma$.  A simple idea is to take $\Umax=\max\{50, 3\hat\sigma\}$.  
\item For the initial guess $\psi_0$ of the mixing density, there are many possibilities.  Here we make a ``non-informative'' choice, taking $\psi_0$ to be a uniform density on $\U$.  One could also consider an ``informative'' choice of $\psi_0$, e.g., a gamma density, truncated to $\U$, with mode at the least squares estimator $\hat\sigma$ of the normal scale $\sigma$.  
\item As discussed previously, to weaken the dependence of the PR estimates on the data ordering, we recommend averaging over 25 data permutations.  These permutations can be selected at random, but it is important that the permutations remain fixed throughout the optimization process.  
\end{itemize} 

To estimate $\vbeta$, we propose the PR maximum likelihood estimator $\hat\vbeta$, the maximizer of $\Lpr(\vbeta)$ or $\lpr(\vbeta)$.  To compute the estimator, one strategy is to use a prepackaged numerical optimization routine.  However, for relatively high-dimensional problems, direct optimization seems to be too costly, so we opt for a more efficient alternative based on the structure of the mixture problem; see Section~\ref{SS:em}.  

Theoretical questions about existence and uniqueness of the maximum PR likelihood estimator are difficult to answer; this is a result of the complicated recursive structure of the PR algorithm.  \citet{mt-prml} make the conjecture that, under some conditions, $\lpr(\vbeta)$ is a concave function of $\vbeta$.  Concavity would guarantee that a unique maximizer of $\lpr$ could be found in practice.  Moreover, concavity could also be used to establish asymptotic consistency of the PR maximum likelihood estimator \citep[e.g.,][]{hjort.pollard.1993}.  A host of examples, including our Figure~\ref{fig:likpic}, support this conjecture, but currently no theory is available; see Section~\ref{S:discuss}.

\subsection{Computation: a hybrid PR--EM algorithm}
\label{SS:em}

The goal is to maximize the PR likelihood $\Lpr(\vbeta)$ or the log-likelihood $\lpr(\vbeta)$.  There is a computational gain that comes from taking advantage of the special structure of the problem.  Along these lines, we present a hybrid PR--EM algorithm for maximizing $\Lpr(\vbeta)$.  The jumping off point here is an alternative interpretation of the scale mixture formulation in \eqref{eq:mixture} in terms of latent scale parameters $U_1,\ldots,U_n$.  Then we have the following trivial identity:
\begin{align*}
\lpr(\vbeta) & = \sum_{i=1}^n \log \int \nm(y_i - \xvec_i \vbeta \mid 0, u^2) \psi_{i-1,\vbeta}(u) \,du \\
& = \sum_{i=1}^n \log \nm(y_i - \xvec_i \vbeta \mid 0, U_i^2) - \sum_{i=1}^n \log\Bigl\{ \frac{\nm(y_i - \xvec_i \vbeta \mid 0, U_i^2)}{f_{i-1,\vbeta}(y_i - \xvec_i \vbeta)} \Bigr\}.
\end{align*}
Since this holds for all $U_1,\ldots,U_n$, it must also hold if we take expectation with respect to some distribution over $U_1,\ldots,U_n$.  Our proposal is to integrate out $U_i$ with respect to the density 
\begin{equation}
\label{eq:posterior}
\psi_{i,\hat\vbeta}^B(u) = \frac{\nm(y_i - \xvec_i \hat\vbeta \mid 0, u^2) \psi_{i-1,\hat\vbeta}(u)}{f_{i-1,\hat\vbeta}(y_i - \xvec_i \hat\vbeta)}, 
\end{equation}
where $\hat\vbeta$ is some estimate.  This is exactly the Bayes posterior density based on ``prior'' $\psi_{i-1,\hat\vbeta}$ and ``data'' $y_i - \xvec_i \hat\vbeta$.  In particular, write 
\begin{align*}
\lpr(\vbeta) = \sum_{i=1}^n \int & \log \nm(y_i - \xvec_i \vbeta \mid 0, u^2) \psi_{i,\hat\vbeta}^B(u) \,du \\
& - \sum_{i=1}^n \int \log\Bigl\{ \frac{\nm(y_i - \xvec_i \vbeta \mid 0, u^2)}{f_{i-1,\vbeta}(y_i - \xvec_i \vbeta)} \Bigr\} \psi_{i,\hat\vbeta}^B(u) \,du.
\end{align*}
Write this as $Q_1(\vbeta \mid \hat\vbeta) + Q_2(\vbeta \mid \hat\vbeta)$.  Then $Q_1(\vbeta \mid \hat\vbeta)$ simplifies to 
\[ Q_1(\vbeta \mid \hat\vbeta) = -\frac12 \sum_{i=1}^n \hat\omega_i (y_i - \xvec_i \vbeta)^2 + \text{constant}, \]
where the weight $\omega_i$, which depends on $\psi_{i-1,\hat\vbeta}$ and $y_i-\xvec_i \hat\vbeta$, is given by 
\begin{equation}
\label{eq:prem.weight}
\hat\omega_i = \int u^{-2} \psi_{i,\hat\vbeta}^B(u) \, du, 
\end{equation}
the expected precision (inverse variance) under the distribution with density in \eqref{eq:posterior}.  Both $\hat\omega_i$ and the constant term depend on $\hat\vbeta$, but not on $\vbeta$.  

We are now ready to state the hybrid PR--EM algorithm.  As with all EM algorithms, we have written the objective function as a sum of two functions, and the idea is that iteratively maximizing $Q_1$ will generate a sequence of parameter values tending to the maximizer of the the original objective function.  This is a desirable approach because maximizing $Q_1(\vbeta \mid \hat\vbeta)$ corresponds to a weighted least squares problem, for which an analytic solution is available.  Justification for the claimed ascent property of PR--EM, which involves some investigation into the $Q_2$ function, is given in Section~\ref{SS:ascent}.  


\begin{PREMalg}
Initialize the algorithm by choosing $\hat\vbeta=\hat\vbeta^{(1)}$ and setting the input $(\U,\psi_0,w_1,\ldots,w_n)$ for the PR portion.  At iteration $t$, do the following steps.  
\begin{description}
\item[\it E-step.] Compute the weights $\hat\omega_1,\ldots,\hat\omega_n$ by running the PR algorithm with the residuals $y_1-\xvec_1 \hat\vbeta^{(t)},\ldots,y_n-\xvec_n \hat\vbeta^{(t)}$ as data. 
\vspace{-2mm}
\item[\it M-step.] Choose $\hat\vbeta^{(t+1)}$ to maximize $Q_1(\vbeta \mid \hat\vbeta^{(t)})$, i.e., $\hat\vbeta^{(t+1)} = (\Xmat^\top \widehat\vOmega \Xmat)^{-1} \Xmat^\top \widehat\vOmega \yvec$, where $\widehat\vOmega$ is a diagonal matrix of the weights $\hat\omega_1,\ldots,\hat\omega_n$. 
\end{description}
Stop when $\|\hat\vbeta^{(t+1)}-\hat\vbeta^{(t)}\|_1 < \delta$, for a specified tolerance $\delta > 0$, and return the corresponding estimates of $\vbeta$ and $\psi$, as well as the weights $\hat\omega_1,\ldots,\hat\omega_n$.  
\end{PREMalg}

In our implementation, we initialize $\hat\vbeta^{(1)}$ at the ordinary least squares estimator, and we take the input $(\U,\psi_0,w_1,\ldots,w_n)$ for the PR portion of the algorithm as discussed in Section~\ref{S:prml.reg}.  The only adjustment required to the general PR algorithm in Section~\ref{SSS:pr.non} is to add a step that calculates the weights $\hat\omega_i$ at each iteration.

\subsection{On the ascent property of PR--EM}
\label{SS:ascent}

To motivate our choice for the density in \eqref{eq:posterior} and to justify our calling this a hybrid PR--EM algorithm, we will give a heuristic argument that the usual EM ascent property holds, i.e., if $Q_1(\vbeta \mid \hat\vbeta) \geq Q_1(\hat\vbeta \mid \hat\vbeta)$, then $\lpr(\vbeta) \geq \lpr(\hat\vbeta)$, at least approximately.   We start by rewriting $Q_2(\vbeta \mid \hat\vbeta)$ as follows:
\[ Q_2(\vbeta \mid \hat\vbeta) = -\sum_{i=1}^n \int \log\Bigl\{ \frac{\nm(y_i - \xvec_i \vbeta \mid 0, u^2)}{f_{i-1,\hat\vbeta}(y_i - \xvec_i \vbeta)} \Bigr\} \psi_{i,\hat\vbeta}^B(u) \,du + n D_n(\vbeta, \hat\vbeta), \]
where 
\[ D_n(\vbeta, \hat\vbeta) = \frac1n\sum_{i=1}^n \log \frac{f_{i-1,\vbeta}(y_i-\xvec_i \vbeta)}{f_{i-1,\hat\vbeta}(y_i - \xvec_i \vbeta)}. \]
The $D_n$ term appears above because we have replaced $f_{i-1,\vbeta}(y_i - \xvec_i \vbeta)$ in the denominator inside the integral with $f_{i-1,\hat\vbeta}(y_i-\xvec_i \vbeta)$; the latter quantity is just the PR estimate $f_{i-1,\hat\vbeta}$ evaluated at $y_i-\xvec_i \vbeta$.  It is easy to see that $D_n(\hat\vbeta,\hat\vbeta)=0$, so 
\[ Q_2(\vbeta \mid \hat\vbeta) - Q_2(\hat\vbeta \mid \hat\vbeta) = \sum_{i=1}^n \int \log\Bigl\{ \frac{\psi_{i,\hat\vbeta}^B(u)}{g_{i,\vbeta,\hat\vbeta}(u)} \Bigr\} \psi_{i,\hat\vbeta}^B(u)\,du + nD_n(\vbeta,\hat\vbeta), \]
where
\[ g_{i,\vbeta,\hat\vbeta}(u) = \frac{\nm(y_i - \xvec_i \vbeta \mid 0, u^2) \psi_{i-1,\hat\vbeta}(u)}{f_{i-1,\hat\vbeta}(y_i-\xvec_i \vbeta)}. \]
The integral on the inside is a Kullback--Leibler divergence and, therefore, is non-negative; it equals zero if and only if $\vbeta=\hat\vbeta$.  Therefore, we have 
\[ \lpr(\vbeta) - \lpr(\hat\vbeta) \geq Q_1(\vbeta \mid \hat\vbeta) - Q_1(\hat\vbeta \mid \hat\vbeta) + n D_n(\vbeta,\hat\vbeta). \]
If we had that $D_n(\vbeta,\hat\vbeta) \geq 0$ with equality if and only if $\vbeta=\hat\vbeta$, then we could conclude that, by choosing $\vbeta$ such that $Q_1(\vbeta \mid \hat\vbeta) > Q_1(\hat\vbeta \mid \hat\vbeta)$, one achieves $\lpr(\vbeta) > \lpr(\hat\vbeta)$. The following heuristics explain why the inequality $D_n(\vbeta,\hat\vbeta) \geq 0$ should hold, at least approximately.  Rewrite $D_n$ as 
\[ D_n(\vbeta,\hat\vbeta) = \frac1n\sum_{i=1}^n \log \frac{f^\star(y_i-\xvec_i \vbeta)}{f_{i-1,\hat\vbeta}(y_i - \xvec_i \vbeta)} - \frac1n\sum_{i=1}^n \log \frac{f^\star(y_i-\xvec_i \vbeta)}{f_{i-1,\vbeta}(y_i - \xvec_i \vbeta)}, \]
where $f^\star$ is the true density of the errors.  If we assume that $\vbeta$ is the true value, then the second term converges, as $n \to \infty$, to the smallest Kullback--Leibler divergence from $f^\star$ over all mixtures of the specified form \citep{mt-prml}.  It is not clear if the  first term will converge or not.  If it does converge, then the limit would also be a Kullback--Leibler divergence and, by definition, cannot be smaller than the limit of the second term so, for large $n$, the difference would be non-negative.  This argument is based on the assumption that $\vbeta$ is the true value.  Therefore, the conclusion that we can reach is that if values of $\vbeta$ at or near the true value will increase $Q_1$ for the given $\hat\vbeta$, which is intuitively quite reasonable, then we can expect that those same values will also increase $\lpr$.  The obstacle to making this heuristic argument rigorous is that a theory of the behavior of the PR estimates for the non-iid case is not yet available; see Section~\ref{S:discuss}.

\subsection{On robustness of PR maximum likelihood}
\label{SS:robustness}

Robustness is an important consideration for all statistical methods; some recent examples of detailed robustness studies include \citet{paula.etal.2012} and \citet{leiva.etal.2014}.  A primary motivation for the flexible scale mixture model \eqref{eq:mixture} for the error distribution is to be able to accommodate ``extreme observations'' that might arise when the true error distribution has heavier-than-normal tails.  Therefore, it is important to discuss in what sense the PR maximum likelihood method is robust to these extremes.   

In general, a model-based method, such as our PR-based method, can be insensitive to extreme data points only if the model in consideration is sufficiently broad.  Our proposed scale mixture model \eqref{eq:mixture} includes many heavy-tailed distributions, including the wide class of exponential power family densities \citep[][Sec.~3.2.1]{box.tiao.1973} among others.  Since our model is broad, we can expect that the PR maximum likelihood method will not be overly sensitive to extremes and, therefore, will be robust in this sense.  The numerical examples in Section~\ref{S:numerical} support this claim. 

A formal theoretical study of the robustness of the PR maximum likelihood estimator is challenging and beyond the scope of this paper.  However, it will be helpful to have some further insights on why the method is insensitive to extreme observations.  For this, recall the weights $\hat\omega_1,\ldots,\hat\omega_n$, defined in \eqref{eq:prem.weight}, produced as a by-product of the PR--EM algorithm in Section~\ref{SS:em}; see, also, Equation~(3) in \citet{lange.little.taylor.1989}.  In particular, the weight $\hat\omega_i$ is the expected value of $U^{-2}$ where $U$ is a positive random variable whose distribution has a density, in \eqref{eq:posterior}, proportional to $\nm(y_i - \xvec_i \hat\vbeta \mid 0, u^2) \psi_{i-1, \hat\vbeta}(u)$.  The claim is that, at PR--EM convergence, a weight will be small if the corresponding observed response is an ``outlier'' and, therefore, based on the weighted least squares representation of the PR maximum likelihood estimator in the M-step, that observation must not be overly influential.  To facilitate this discussion, we make the simplifying assumption that the potentially extreme observation in question is $y_n$; the averaging over permutations implies that order of the data is (mostly) irrelevant in the model fitting so this  is essentially without loss of generality.  Suppose that $|y_n - \xvec_n \hat\vbeta|$ is large, so that the normal density factor, $\nm(y_n - \xvec_n \hat\vbeta \mid 0, u^2)$, in the density above will be large only when $u$ is large.  Naturally, the extent to which the observation $y_n$ will be down-weighted depends on the sample size $n$.  Suppose first that $n$ is small.  Then the mixing density factor, $\psi_{n-1,\hat\vbeta}(u)$, will be relatively close to the initial value, $\psi_0(u)$, so the normal density factor will be dominant.  Therefore, in this case, the weight $\hat\omega_n$ will be close to zero so $y_n$ will not be an influential observation.  Now suppose that $n$ is large.  In this case, the mixing density factor will likely be small for large $u$---see Figures~\ref{fig:phones}(c) and \ref{fig:hbk}(d)---so the the large normal density factor may be dominated by the small mixing density factor.  Therefore, the weight $\hat\omega_n$ assigned to an equally extreme $y_n$ may not be particularly small.  This shows that PR--EM's decision on how to weight observations is based not just on the size of the residual but also on the sample size, among other things.  This is desirable since classifying an observation as ``extreme'' should be done relative to the available sample.   



\section{Numerical results}
\label{S:numerical}

\subsection{Methods}
\label{SS:methods}

For the numerical results in this section, we compare our hybrid PR--EM method with the following methods for robust regression; our computations are carried out using the statistical software R \citep{Rmanual}.   
\begin{itemize}
\item[\rm LS.] Ordinary least squares, with the R function {\tt lm};
\item[\rm RLS.] Robust least squares, with the default settings of the R function {\tt rlm};
\item[\rm ML.] Maximum likelihood based on a Student-t error distribution, with $\text{df}=4$, using iteratively re-weighted least squares via {\tt lm};
\item[\rm $L_1$.] Least absolute error regression using the default settings of the R function {\tt rq} in the {\tt quantreg} package \citep{Rquantreg}.
\end{itemize}
The choice of $\text{df=4}$ in ML is based on \citet{lange.little.taylor.1989}, \citet{lange.sinsheimer.1993}, \citet[][Sec.~5.2]{brazzale.davison.reid.2007}, and \citet{barros.etal.2009}.  R code to implement our proposed method, denoted by PREM, is available at \url{www.math.uic.edu/~rgmartin}.

\subsection{Real data analysis}
\label{SS:real}

\begin{figure}[t]
\begin{center}
\subfigure[Scatterplot with fitted regression lines]{\scalebox{0.5}{\includegraphics{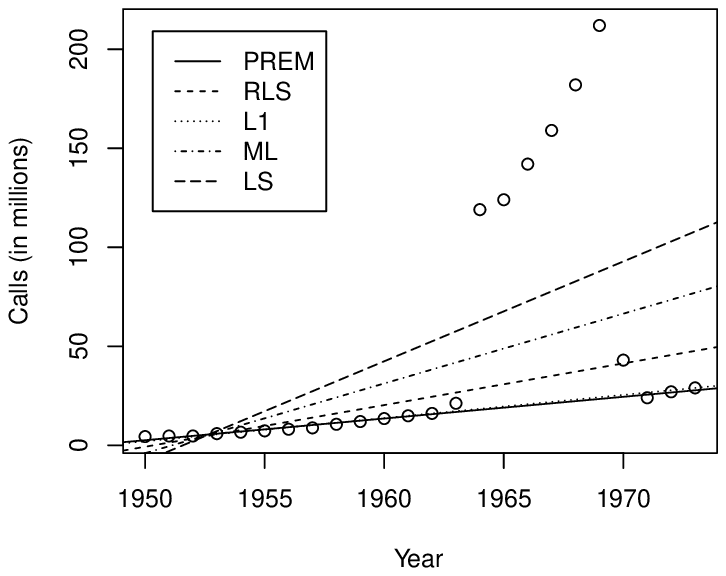}}}
\subfigure[PREM weights $\hat\omega_i$ in \eqref{eq:prem.weight}]{\scalebox{0.5}{\includegraphics{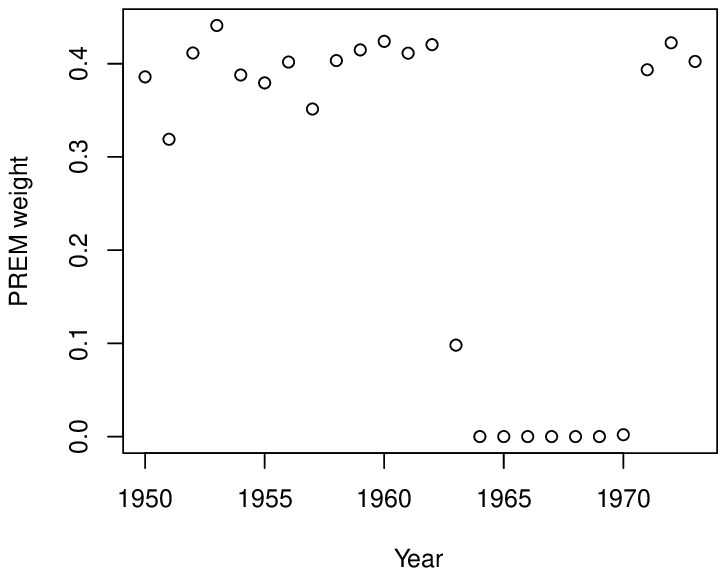}}}
\subfigure[PREM mixing density estimate]{\scalebox{0.5}{\includegraphics{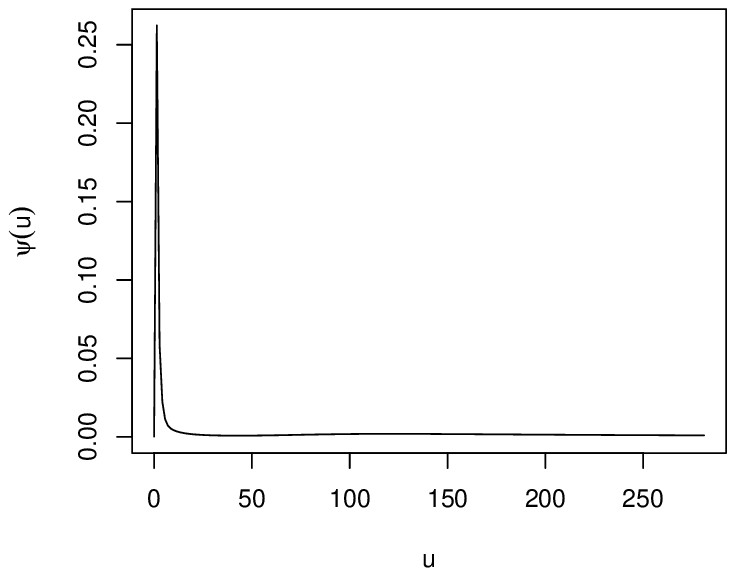}}}
\subfigure[PREM likelihood path]{\scalebox{0.5}{\includegraphics{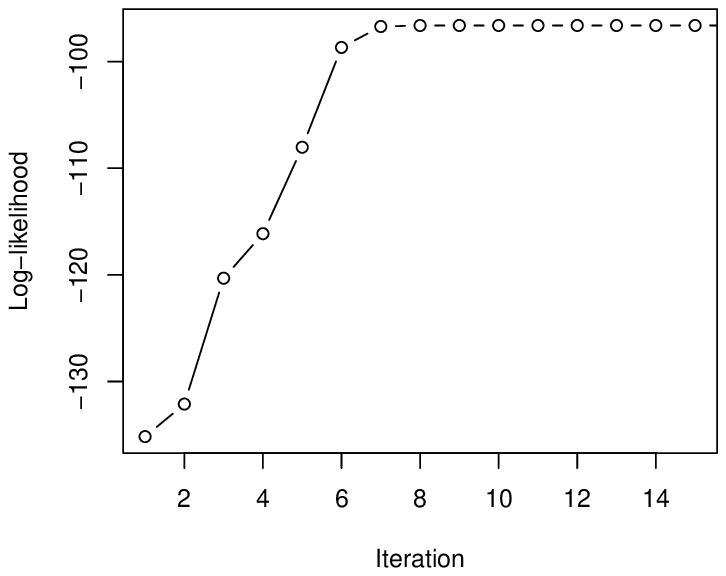}}}
\end{center}
\caption{Results for the Belgian phone call data in Example~\ref{ex:phones}.}
\label{fig:phones}
\end{figure}

\begin{ex}
\label{ex:phones}
Consider a simple linear regression problem, where the predictor variable is the year, ranging from 1950 to 1973, and the response variable is the number of international phone calls from Belgium each year; so $n=24$.  These data, available in the R software MASS library under the name {\tt phones}, provide a classic example for robust regression \citep{rousseeuw.leroy.1987}.  The scatterplot in Figure~\ref{fig:phones}(a) immediately reveals the presence of several vertical outliers.  The fitted lines for four methods are shown overlaid the scatter plot.  An immediate conclusion is that the PREM estimate is not influenced by the outliers at all, while the vertical outliers make the other methods' estimates (except $L_1$) too steep to fit the data at the later dates.  

For more on PREM, we give three additional displays.  First, in Figure~\ref{fig:phones}(b), is a plot of the weights $\hat\omega_i$ in \eqref{eq:prem.weight}.  The observations with weights near 0 are exactly those apparent outliers in Figure~\ref{fig:phones}(a).  That these observations are assigned nearly 0 weight explains why they had essentially no influence on the fitting of the regression line.  Figure~\ref{fig:phones}(c) shows a plot of the PREM mixing density $\psi$.  Most of the mass is close to 0, consistent with the fact that the fluctuations around the fitted line is minimal, but there is a wide, almost imperceptible bump near $u=100$ which is accounting for the vertical outliers.  Figure~\ref{fig:phones}(d) displays the PR log-marginal likelihood path versus PR--EM iterations, and the monotonicity of the path is confirmed.  
\end{ex}

\begin{figure}[t]
\begin{center}
\subfigure[Plot of LS residuals]{\scalebox{0.5}{\includegraphics{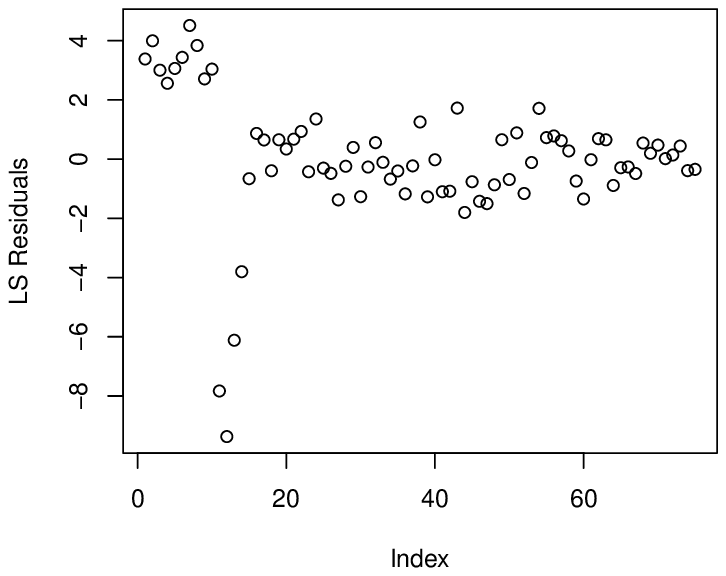}}}
\subfigure[Quantile plot of LS residuals]{\scalebox{0.5}{\includegraphics{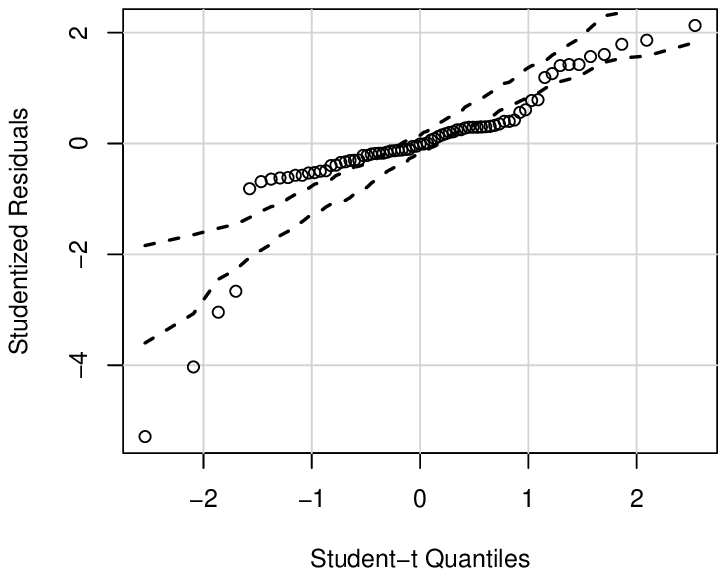}}}
\subfigure[PREM residuals]{\scalebox{0.5}{\includegraphics{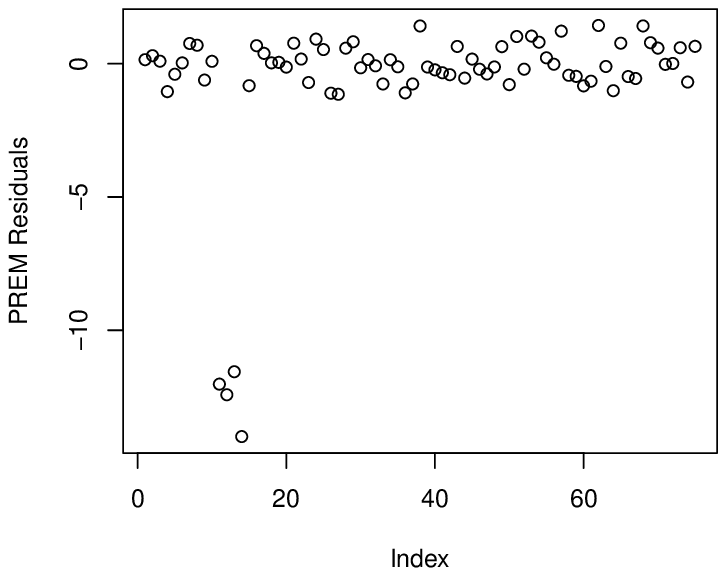}}}
\subfigure[PREM mixing density estimate]{\scalebox{0.5}{\includegraphics{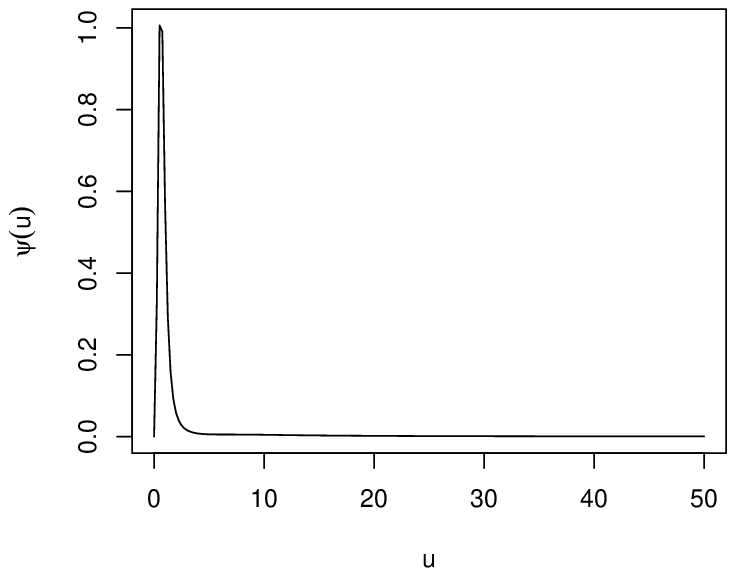}}}
\end{center}
\caption{Results for the Hawkins--Bradu--Kass data in Example~\ref{ex:hbk}.}
\label{fig:hbk}
\end{figure}

\begin{ex}
\label{ex:hbk}
Here we consider the data presented in \citet{hawkins.bradu.kass.1984}, consisting of $n=75$ observations and three predictor variables.  This example is considered as a benchmark for outlier detection methods.  The first ten observations are regression outliers, i.e., deviations from the overall linear pattern, and the next four observations are $\xvec$-outliers, or leverage points.  A plot of the LS residuals versus observation number is displayed in Figure~\ref{fig:hbk}(a), which demonstrates the characteristics of these first 14 troublesome observations.  A quantile plot of the least squares residuals, with simulated envelope \citep{atkinson1985}, is shown in Figure~\ref{fig:hbk}(b), which reveals the non-normality in the residuals.  The PREM weights (not displayed) assigned to the four leverage points are effectively zero, so these points have no influence to the PREM fit.  Figure~\ref{fig:hbk}(c) shows the PREM residuals, and it is clear that the PREM fit is good for all points except the four leverage points assigned weight near zero.  The estimated mixing density is displayed in Figure~\ref{fig:hbk}(d), and it concentrates its mass in a small interval around the least squares estimator $\hat\sigma=2.25$.  This example shows that the PR--EM approach both removes the $\xvec$-outliers in the model-fitting step by assigning them negligible weight and accommodates the regression outliers with a flexible model for the errors.  
\end{ex}

\begin{figure}[t]
\begin{center}
\subfigure[Quantile plot of LS residuals]{\scalebox{0.5}{\includegraphics{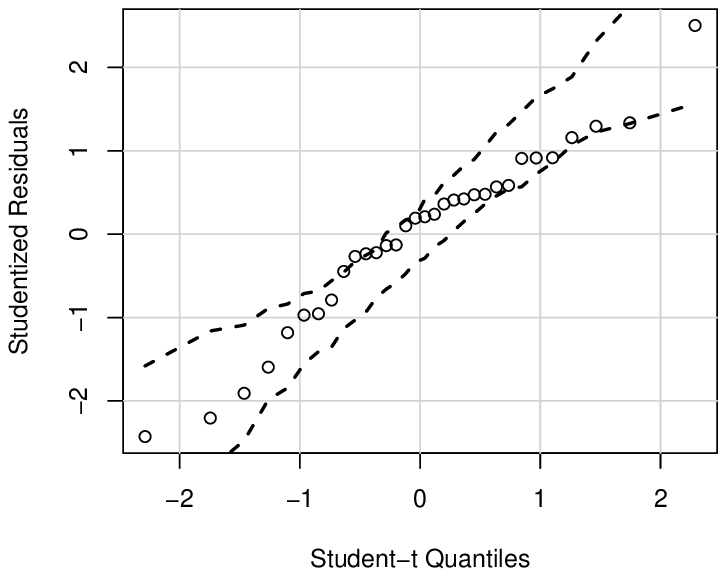}}}
\subfigure[95\% confidence intervals]{\scalebox{0.5}{\includegraphics{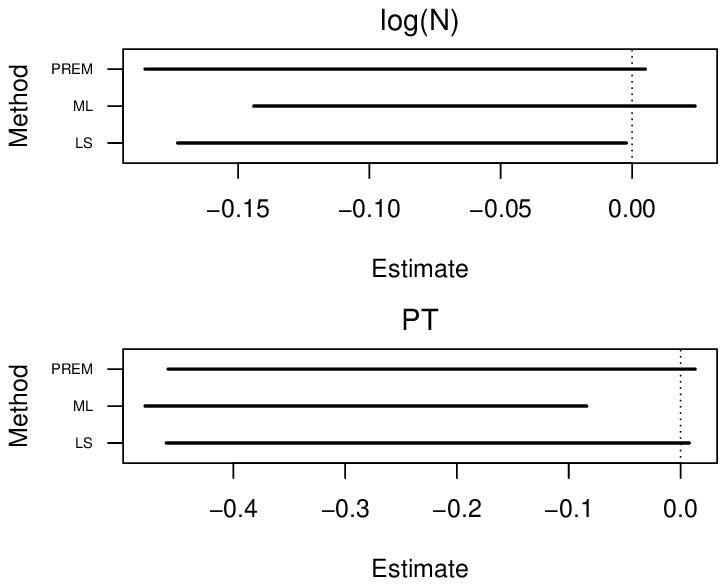}}}
\end{center}
\caption{Results for the nuclear plant data in Example~\ref{ex:nuclear}.}
\label{fig:nuclear}
\end{figure}

\begin{ex}
\label{ex:nuclear}
In the first two examples, the presence of outliers and/or non-normality was clear; in this example, whether there is a departure from the standard Gaussian linear regression assumption is less clear.  \citet[][Example~G]{cox.snell.1981} present an example involving data on the construction of $n=32$ light water reactor plants, and the mean log-cost to construct a nuclear reactor is modeled as a linear function of several predictor variables; see, also, \citet{davison.hinkley.1997}, \citet{brazzale.davison.reid.2007}, and \citet{koller.stahel.2011}.  After an initial variable screening, a model containing six predictor variables is considered.  Two of these six predictor variables, namely {\tt log(N)} and {\tt PT}, which denote the number of nuclear power plants constructed by each architect--engineer (on the log scale) and an indicator for those plants with partial turnkey guarantees, respectively, are of primary interest here.  In particular, these two variables are only marginally significant based on standard regression techniques, so one could ask whether the significance of these two variables is sensitive to the choice of error distribution.  

Figure~\ref{fig:nuclear}(a) shows a quantile plot of the studentized residuals from a least squares fit, and this suggests a possibly heavier-than-normal tailed error distribution.  This motivates \citet[][Sec.~5.2]{brazzale.davison.reid.2007} to employ the ML method described in Section~\ref{SS:methods}. Figure~\ref{fig:nuclear}(b) plots 95\% confidence intervals for the slope coefficients for {\tt log(N)} and {\tt PT} based on the usual LS distribution theory, the first-order asymptotic normality of ML, and the following method for PREM.  Since $\lpr(\vbeta)$ is a sort of log-likelihood, \citet{mt-prml} suggest that it can be used to construct confidence intervals in usual way.  That is, let $J(\hat\vbeta)$ denote the inverse of the Hessian matrix for $-\lpr(\vbeta)$ at $\vbeta=\hat\vbeta$, the maximizer.  Then a nominal 95\% confidence interval for $\beta_j$ is $\hat\beta_j \pm 1.96\{J(\hat\vbeta)_{jj}\}^{1/2}$.  In this case, the estimated coefficients for PREM and LS are almost indistinguishable, so we expect the confidence intervals to have roughly the same center.  That the PREM confidence intervals are a bit longer is also to be expected since we are fitting a semiparametric model.  However, Figure~\ref{fig:nuclear}(a) suggests that the normal error model is reasonable, so it is promising that the PREM intervals are not too much longer than the LS intervals.  That is, PREM does not substantially over-fit when a normal model is reasonable.  Moreover, when normality is questionable, it may be more reasonable to enlarge the model space, as PREM does, rather than change the model.  When the model space is enlarged, inference should be more conservative, so we argue that the conclusions based on the PREM analysis might be more reasonable than those in \citet{brazzale.davison.reid.2007} and \citet{koller.stahel.2011} which are more aggressive and conclude that {\tt PT} is significant.   
\end{ex}

\subsection{Simulations}

This section provides simulation results to compare the performance of PREM with that of the competitors listed in Section~\ref{SS:methods}, under a variety of error distributions.  We consider six error distributions: the two extreme exponential power distributions \citep{west1987}, namely, the standard normal and the standard Laplace; Student-t distributions with 1 and 2 degrees of freedom, respectively; and two non-standard normal scale mixtures, one with respect to a standard exponential distribution, denoted by {\sf N--Exp}, and the other with respect to a uniform distribution supported on $(0,7)$, denoted by {\sf N--Unif}.  Both of the latter two distributions are of the general form of our model, but our support $\U$ for the mixing density $\psi$ is misspecified in both cases.  Also, {\sf N--Exp} has slightly heavier tails than the Laplace.  These examples are far from exhaustive, but they do demonstrate that the hybrid PR--EM algorithm, with its flexible semiparametric model, is both fast and accurate compared to its competitors in a range of problems.  

\begin{scenario}
\label{ex:low.dim}
In this case, we consider a regression with two predictor variables.  The two predictor variables are taken to be independent standard normal samples, and the true parameter is $\vbeta=\mathbf{1}_3$.  Table~\ref{tab:mse1} gives the empirical mean square error of the estimators over 100 replications in each configuration.  PREM is the only semiparametric estimator, and though it is not the best performer in all cases, it is not dominated by any other method. 
\end{scenario}

\begin{table}
\begin{center}
\begin{tabular}{ccccccc}
\hline
$f$ & LS & RLS & ML & $L_1$ & PREM \\
\hline
${\sf N}$ & 0.029 & 0.031 & 0.033 & 0.047 & 0.037 \\
${\sf Laplace}$ & 0.069 & 0.050 & 0.048 & 0.049 & 0.048 \\
${\sf t}_1$ & 82.2 & 0.127 & 0.117 & 0.0924 & 0.099 \\
${\sf t}_2$ & 0.281 & 0.065 & 0.060 & 0.068 & 0.067 \\ 
{\sf N}--{\sf Exp} & 0.060 & 0.019 & 0.018 & 0.007 & 0.008 \\
{\sf N}--{\sf Unif} & 0.500 & 0.317 & 0.292 & 0.212 & 0.229 \\
\hline
\end{tabular}
\end{center}
\caption{Empirical mean squared error for the indicated method and distribution in the simulation described in Scenario~\ref{ex:low.dim}.}
\label{tab:mse1}
\end{table}

\begin{scenario}
\label{ex:higher.dim}
Here we consider a higher dimensional version of the simulation described in Example~\ref{ex:low.dim}.  This time we take $p=10$ predictor variables, including the intercept, $\vbeta=\mathbf{1}_{10}$, and, except for the intercept term, introduce some dependence in the predictor variables by sampling each case from a $(p-1)$-dimensional normal distribution with mean zero and autoregression covariance structure, with correlation parameter $\rho=0.5$.  The same empirical mean square errors, as in Table~\ref{tab:mse1}, are presented in Table~\ref{tab:mse2}.  Again, the semiparametric PREM is competitive with existing parametric methods.  
\end{scenario}

\begin{table}
\begin{center}
\begin{tabular}{ccccccc}
\hline
$f$ & LS & RLS & ML & $L_1$ & PREM \\
\hline
${\sf N}$ & 0.171 & 0.181 & 0.185 & 0.264 & 0.255 \\
${\sf Laplace}$ & 0.388 & 0.290 & 0.280 & 0.311 & 0.336 \\
${\sf t}_1$ & 608 & 0.837 & 0.791 & 0.665 & 0.699 \\
${\sf t}_2$ & 1.412 & 0.375 & 0.347 & 0.373 & 0.394 \\ 
{\sf N}--{\sf Exp} & 0.359 & 0.127 & 0.123 & 0.079 & 0.066 \\
{\sf N}--{\sf Unif} & 2.770 & 2.034 & 1.906 & 1.603 & 1.865 \\
\hline
\end{tabular}
\end{center}
\caption{Empirical mean squared error for the indicated method and distribution in the simulation described in Scenario~\ref{ex:higher.dim}.}
\label{tab:mse2}
\end{table}

\section{Discussion}
\label{S:discuss}

This paper proposes a flexible semiparametric model in which the error distribution is taken to be a general scale mixture of normals with unknown mixing distribution.  We estimate the regression coefficients $\vbeta$ by maximizing the PR-based likelihood function $\Lpr(\vbeta)$ and, for this purpose, we have developed a hybrid PR--EM algorithm based on the scale mixture of normals model for the error terms.  As a by-product of the hybrid algorithm, scores are produced for each observation which can be used for outlier detection and also justify the robustness of the estimator.  

The PR method in general has proved to be a useful tool in a variety of problems.  However, its complicated recursive structure makes it difficult to analyze theoretically.  For this reason, there remains a number of interesting open questions regarding its behavior, both asymptotics and finite samples.  In particular, as we discussed above, concavity of the log-likelihood $\lpr(\vbeta)$ is important for various questions about the PR estimator, i.e., existence of the estimator in finite samples, and asymptotic consistency.  Theoretical study of PR has been so far limited to iid models, but, as explained in Section~\ref{SS:em}, there is a need for further work in the independent non-iid case.  We hope that the work here will motivate further studies of PR both in and beyond the iid setup.  


\section*{Acknowledgments}

The authors are grateful to the Editor, Associate Editor, and three referees for helpful comments and suggestions, and to Byungtae Seo for providing the slides from his talk at the 2012 Joint Statistical Meetings.

\bibliographystyle{apalike}
\bibliography{/Users/rgmartin/Dropbox/Research/mybib}

\begin{thebibliography}{}

\bibitem[Andrews and Mallows, 1974]{andrews.mallows.1974}
Andrews, D.~F. and Mallows, C.~L. (1974).
\newblock Scale mixtures of normal distributions.
\newblock {\em J. Roy. Statist. Soc. Ser. B}, 36:99--102.

\bibitem[Atkinson, 1985]{atkinson1985}
Atkinson, A.~C. (1985).
\newblock {\em Plots, Transformations, and Regression}.
\newblock Oxford Univ. Press, Oxford.

\bibitem[Barros et~al., 2009]{barros.etal.2009}
Barros, M., Paula, G.~A., and Leiva, V. (2009).
\newblock An {R} implementation for generalized {B}irnbaum-{S}aunders
  distributions.
\newblock {\em Comput. Statist. Data Anal.}, 53(4):1511--1528.

\bibitem[Box and Tiao, 1973]{box.tiao.1973}
Box, G. E.~P. and Tiao, G.~C. (1973).
\newblock {\em Bayesian Inference in Statistical Analysis}.
\newblock Addison-Wesley Publishing Co., Reading, Mass.-London-Don Mills, Ont.

\bibitem[Brazzale et~al., 2007]{brazzale.davison.reid.2007}
Brazzale, A.~R., Davison, A.~C., and Reid, N. (2007).
\newblock {\em Applied Asymptotics: Case Studies in Small-Sample Statistics}.
\newblock Cambridge University Press, Cambridge.

\bibitem[Carvalho et~al., 2010]{carvalho.ba.2010}
Carvalho, C.~M., Lopez, H.~F., Polson, N.~G., and Taddy, M.~A. (2010).
\newblock Particle learning for general mixtures.
\newblock {\em Bayesian Anal.}, 5:709--740.

\bibitem[Chib, 1995]{chib1995}
Chib, S. (1995).
\newblock Marginal likelihood from the {G}ibbs output.
\newblock {\em J. Amer. Statist. Assoc.}, 90(432):1313--1321.

\bibitem[Cox and Snell, 1981]{cox.snell.1981}
Cox, D.~R. and Snell, E.~J. (1981).
\newblock {\em Applied Statistics}.
\newblock Chapman \& Hall, London.

\bibitem[Davison and Hinkley, 1997]{davison.hinkley.1997}
Davison, A.~C. and Hinkley, D.~V. (1997).
\newblock {\em Bootstrap Methods and their Application}, volume~1.
\newblock Cambridge University Press, Cambridge.

\bibitem[Dempster et~al., 1977]{dlr}
Dempster, A., Laird, N., and Rubin, D. (1977).
\newblock Maximum-likelihood from incomplete data via the {EM} algorithm (with
  discussion).
\newblock {\em J. Roy. Statist. Soc. Ser. B}, 39(1):1--38.

\bibitem[Escobar and West, 1995]{escobar.west.1995}
Escobar, M.~D. and West, M. (1995).
\newblock Bayesian density estimation and inference using mixtures.
\newblock {\em J. Amer. Statist. Assoc.}, 90(430):577--588.

\bibitem[Ferguson, 1973]{ferguson1973}
Ferguson, T.~S. (1973).
\newblock A {B}ayesian analysis of some nonparametric problems.
\newblock {\em Ann. Statist.}, 1:209--230.

\bibitem[Ghosh and Tokdar, 2006]{ghoshtokdar}
Ghosh, J.~K. and Tokdar, S.~T. (2006).
\newblock Convergence and consistency of {N}ewton's algorithm for estimating
  mixing distribution.
\newblock In Fan, J. and Koul, H., editors, {\em Frontiers in Statistics},
  pages 429--443. Imp. Coll. Press, London.

\bibitem[Hawkins et~al., 1984]{hawkins.bradu.kass.1984}
Hawkins, D.~M., Bradu, D., and Kass, G.~V. (1984).
\newblock Location of several outliers in multiple-regression data using
  elemental sets.
\newblock {\em Technometrics}, 26(3):197--208.

\bibitem[Hjort and Pollard, 1993]{hjort.pollard.1993}
Hjort, N.~L. and Pollard, D. (1993).
\newblock Asymptotics for minimisers of convex processes.
\newblock Unpublished manuscript,
  \url{http://www.stat.yale.edu/~pollard/Papers/convex.pdf}.

\bibitem[Huber, 1973]{huber1973}
Huber, P.~J. (1973).
\newblock Robust regression: asymptotics, conjectures and {M}onte {C}arlo.
\newblock {\em Ann. Statist.}, 1:799--821.

\bibitem[Huber, 1981]{huber1981}
Huber, P.~J. (1981).
\newblock {\em Robust Statistics}.
\newblock John Wiley \& Sons Inc., New York.
\newblock {W}iley Series in Probability and Mathematical Statistics.

\bibitem[Jara et~al., 2011]{dppackage}
Jara, A., Hanson, T., Quintana, F., M\"uller, P., and Rosner, G. (2011).
\newblock {DPpackage}: Bayesian semi- and nonparametric modeling in {R}.
\newblock {\em J. Statist. Softw.}, 40(5):1--30.

\bibitem[Koenker, 2013]{Rquantreg}
Koenker, R. (2013).
\newblock {\em quantreg: Quantile Regression}.
\newblock R package version 5.05.

\bibitem[Koller and Stahel, 2011]{koller.stahel.2011}
Koller, M. and Stahel, W.~A. (2011).
\newblock Sharpening {W}ald-type inference in robust regression for small
  samples.
\newblock {\em Comput. Statist. Data Anal.}, 55(8):2504--2515.

\bibitem[Lange and Sinsheimer, 1993]{lange.sinsheimer.1993}
Lange, K. and Sinsheimer, J.~S. (1993).
\newblock Normal/independent distributions and their applications in robust
  regression.
\newblock {\em J. Comput. Graph. Statist.}, 2(2):175--198.

\bibitem[Lange et~al., 1989]{lange.little.taylor.1989}
Lange, K.~L., Little, R. J.~A., and Taylor, J. M.~G. (1989).
\newblock Robust statistical modeling using the {$t$} distribution.
\newblock {\em J. Amer. Statist. Assoc.}, 84(408):881--896.

\bibitem[Leiva et~al., 2014]{leiva.etal.2014}
Leiva, V., Saulo, H., Le{\~a}o, J., and Marchant, C. (2014).
\newblock A family of autoregressive conditional duration models applied to
  financial data.
\newblock {\em Comput. Statist. Data Anal.}, 79:175--191.

\bibitem[Liu, 1996]{cliu.1996}
Liu, C. (1996).
\newblock Bayesian robust multivariate linear regression with incomplete data.
\newblock {\em J. Amer. Statist. Assoc.}, 91(435):1219--1227.

\bibitem[Lo, 1984]{lo1984}
Lo, A.~Y. (1984).
\newblock On a class of {B}ayesian nonparametric estimates. {I}. {D}ensity
  estimates.
\newblock {\em Ann. Statist.}, 12(1):351--357.

\bibitem[MacEachern and M{\"u}ller, 1998]{maceachern.muller.1998}
MacEachern, S. and M{\"u}ller, P. (1998).
\newblock Estimating mixture of {D}irichlet process models.
\newblock {\em J. Comput. Graph. Statist.}, 7:223--238.

\bibitem[Martin, 2013]{prml-finite}
Martin, R. (2013).
\newblock An approximate {B}ayesian marginal likelihood approach for estimating
  finite mixtures.
\newblock {\em Comm. Statist. Simulation Comput.}, 42(7):1533--1548.

\bibitem[Martin and Ghosh, 2008]{martinghosh}
Martin, R. and Ghosh, J.~K. (2008).
\newblock Stochastic approximation and {N}ewton's estimate of a mixing
  distribution.
\newblock {\em Statist. Sci.}, 23(3):365--382.

\bibitem[Martin and Tokdar, 2009]{mt-rate}
Martin, R. and Tokdar, S.~T. (2009).
\newblock Asymptotic properties of predictive recursion: robustness and rate of
  convergence.
\newblock {\em Electron. J. Stat.}, 3:1455--1472.

\bibitem[Martin and Tokdar, 2011]{mt-prml}
Martin, R. and Tokdar, S.~T. (2011).
\newblock Semiparametric inference in mixture models with predictive recursion
  marginal likelihood.
\newblock {\em Biometrika}, 98(3):567--582.

\bibitem[Martin and Tokdar, 2012]{mt-test}
Martin, R. and Tokdar, S.~T. (2012).
\newblock A nonparametric empirical {B}ayes framework for large-scale multiple
  testing.
\newblock {\em Biostatistics}, 13(3):427--439.

\bibitem[M{\"u}ller and Quintana, 2004]{muller.quintana.2004}
M{\"u}ller, P. and Quintana, F.~A. (2004).
\newblock Nonparametric {B}ayesian data analysis.
\newblock {\em Statist. Sci.}, 19(1):95--110.

\bibitem[Neal, 2000]{neal2000}
Neal, R.~M. (2000).
\newblock Markov chain sampling methods for {D}irichlet process mixture models.
\newblock {\em J. Comput. Graph. Statist.}, 9(2):249--265.

\bibitem[Newton, 2002]{newton02}
Newton, M.~A. (2002).
\newblock On a nonparametric recursive estimator of the mixing distribution.
\newblock {\em Sankhy\=a Ser. A}, 64(2):306--322.

\bibitem[Newton et~al., 1998]{nqz}
Newton, M.~A., Quintana, F.~A., and Zhang, Y. (1998).
\newblock Nonparametric {B}ayes methods using predictive updating.
\newblock In Dey, D., M{\"u}ller, P., and Sinha, D., editors, {\em Practical
  nonparametric and semiparametric Bayesian statistics}, volume 133 of {\em
  Lecture Notes in Statist.}, pages 45--61. Springer, New York.

\bibitem[Paula et~al., 2012]{paula.etal.2012}
Paula, G.~A., Leiva, V., Barros, M., and Liu, S. (2012).
\newblock Robust statistical modeling using the {B}irnbaum-{S}aunders-{$t$}
  distribution applied to insurance.
\newblock {\em Appl. Stoch. Models Bus. Ind.}, 28(1):16--34.

\bibitem[Pinheiro et~al., 2001]{pinheiro.liu.wu.2001}
Pinheiro, J.~C., Liu, C., and Wu, Y.~N. (2001).
\newblock Efficient algorithms for robust estimation in linear mixed-effects
  models using the multivariate {$t$} distribution.
\newblock {\em J. Comput. Graph. Statist.}, 10(2):249--276.

\bibitem[{R Core Team}, 2013]{Rmanual}
{R Core Team} (2013).
\newblock {\em R: A Language and Environment for Statistical Computing}.
\newblock R Foundation for Statistical Computing, Vienna, Austria.

\bibitem[Rogers and Tukey, 1972]{rogers.tukey.1972}
Rogers, W.~H. and Tukey, J.~W. (1972).
\newblock Understanding some long-tailed symmetrical distributions.
\newblock {\em Statistica Neerlandica}, 26(3):211--226.

\bibitem[Rousseeuw and Leroy, 1987]{rousseeuw.leroy.1987}
Rousseeuw, P.~J. and Leroy, A.~M. (1987).
\newblock {\em Robust Regression and Outlier Detection}.
\newblock Wiley Series in Probability and Mathematical Statistics: Applied
  Probability and Statistics. John Wiley \& Sons Inc., New York.

\bibitem[Ryan, 2009]{ryan2009}
Ryan, T.~P. (2009).
\newblock {\em Modern Regression Methods}.
\newblock Wiley Series in Probability and Statistics. John Wiley \& Sons Inc.,
  Hoboken, NJ, second edition.

\bibitem[Tokdar et~al., 2009]{tmg}
Tokdar, S.~T., Martin, R., and Ghosh, J.~K. (2009).
\newblock Consistency of a recursive estimate of mixing distributions.
\newblock {\em Ann. Statist.}, 37(5A):2502--2522.

\bibitem[Wang, 2007]{wang}
Wang, Y. (2007).
\newblock On fast computation of the non-parametric maximum likelihood estimate
  of a mixing distribution.
\newblock {\em J. R. Stat. Soc. Ser. B}, 69(2):185--198.

\bibitem[West, 1987]{west1987}
West, M. (1987).
\newblock On scale mixtures of normal distributions.
\newblock {\em Biometrika}, 74(3):646--648.

\end{thebibliography}

\end{document}